\documentclass[
  prd,
  aps,
  reprint,
  superscriptaddress,
  nofootinbib,
  floatfix,
  amsfonts,
  amssymb,
  amsmath,
  showkeys
]{revtex4-2}

\usepackage{graphicx}
\graphicspath{{figs/}{figs/tikz/}}
\usepackage{tikz}
\usetikzlibrary{arrows.meta,positioning,shapes.geometric,fit,backgrounds,calc,decorations.pathreplacing}
\usepackage[mode=tex]{standalone}
\usepackage{xcolor}
\usepackage{xurl}
\usepackage[colorlinks=true,
            linkcolor=blue,
            citecolor=blue,
            urlcolor=blue]{hyperref}
\usepackage{bm}
\usepackage{booktabs}
\usepackage{siunitx}
\usepackage{mathtools}
\usepackage{xspace}
\usepackage{hyperref}

\newcommand{\jax}{\textsc{JAX}\xspace}
\newcommand{\coppuccino}{\textsc{coppuccino}\xspace}
\newcommand{\flowjax}{\textsc{flowjax}\xspace}
\newcommand{\discovery}{\textsc{discovery}\xspace}

\newcommand{\blackjaxns}{\textsc{blackjax-ns}\xspace}
\newcommand{\logten}{\log_{10}}
\newcommand{\BFPG}{\mathrm{BF}^{\mathrm{PL}}_{\mathrm{Gauss}}}
\newcommand{\bs}{\boldsymbol}

\begin{document}

\title{Fast and Flow-rious: Gravitational Wave Background Bayesian Model Comparison using Normalizing Flows and Nested Sampling}

\author{David C.\ Wright}
\email{wrightd2@oregonstate.edu}
\affiliation{Department of Physics, Oregon State University, Corvallis, Oregon 97331, USA}

\author{Aaron D.\ Johnson}
\affiliation{NASA Marshall Space Flight Center, Huntsville, Alabama 35812, USA}

\author{Jeffrey S.\ Hazboun}
\affiliation{Department of Physics, Oregon State University, Corvallis, Oregon 97331, USA}

\author{William G.\ Lamb}
\affiliation{Department of Physics and Astronomy, Vanderbilt University, 2301 Vanderbilt Place, Nashville, Tennessee 37235, USA}

\date{\today}

\begin{abstract}
In 2023, pulsar timing array (PTA) collaborations around the world announced evidence for a nanohertz gravitational-wave background (GWB).
Beyond increasing the detection significance, the next major goal is to \emph{characterize} the GWB and identify its source(s).
The typical model-comparison metric used in PTA analyses is the Bayes factor (BF).
However, BF computation methods used in the PTA literature, such as product-space sampling and thermodynamic integration, are computationally expensive or inefficient if the models compared are dissimilar or non-nested. 
In this work, we introduce a PTA model-comparison method using a combination of normalizing flows, nested sampling, and a PTA likelihood that is marginalized over all non-GWB parameters.
The flows are trained using a newly-developed \textsc{Python} package \coppuccino{}, which learns the data's \textit{copula}: the dependence structure after transforming to uniform marginals.
We validate the method on both analytic targets and simulated PTA data, recovering the analytic evidence within error bars and achieving an area-under-the-curve (AUC) score of 0.996 when treating the BFs as binary classifiers of the injected models.
The result is a method to perform Bayesian model comparison of arbitrary GWB spectral models in minutes on consumer hardware. 
\end{abstract}

\keywords{gravitational waves; pulsar timing arrays; gravitational-wave background; Bayesian model comparison; normalizing flows; nested sampling}
\maketitle

\section{Introduction}\label{sec:intro}
In recent years, pulsar timing array (PTA) collaborations have reported evidence for a nanohertz gravitational-wave background (GWB) at the $\sim 3 \sigma$~level \cite{NANOGrav:2023gor,reardonSearchIsotropicGravitationalwave2023,antoniadisSecondDataRelease2023,agazieComparingRecentPulsar2024,miles2025meerkat}.
PTA science is now primarily focused on increasing the significance of the measurement \cite{Yu:2025dor,agazieComparingRecentPulsar2024,Agarwal:2026dlv} and characterizing the GWB \cite{NANOGrav:2023hfp,NANOGrav:2023hvm,Agazie:2026tui}.

An observed nanohertz-frequency GWB may be generated by a population of supermassive black hole binaries \cite{rajagopalUltralowfrequencyGravitational1995,jaffeGravitationalWavesProbe2003,wyitheLowFrequencyGravitational2003,sesanaStochasticGravitationalwaveBackground2008} or by ``exotic'' cosmological processes, such as inflationary gravitational waves \cite{laskyGravitationalwaveCosmologyAcross2016,guzzettiGravitationalWavesInflation2016} and phase transitions in the early universe \cite{kamionkowskiGravitationalRadiationFirstorder1994,capriniCosmologicalBackgroundsGravitational2018}.
Determining the source of the GWB is the next priority for pulsar timing array science.
The statistical characteristics of an astrophysical and cosmological GWB differ with respect to their Gaussianity, isotropy, and polarization, which may be measurable by pulsar timing arrays to distinguish between these two origins \cite{Lamb:2025niq,Lamb:2024gbh,sato-politoExploringSpectrumStochastic2024,lemkeDetectingGravitationalWave2025,gardinerBackgroundGravitationalwaveAnisotropy2024}.
However, in this study, we limit ourselves to the fiducial GWB model---a stationary, isotropic, unpolarized, zero-mean Gaussian Process (GP) \cite{vanHaasteren:2014qva,Lentati:2012xb}.
Our methods can be easily extended beyond these minimal scenarios.

The GP is defined by a power spectral density (PSD), which characterizes how the GWB's power is distributed across frequencies.
We call this distribution the ``shape'' of the PSD, which is typically unique to the GWB source.
For example, an astrophysical GWB is expected to generate a double-powerlaw PSD spectrum, where the break (or ``turnover'') frequency between each powerlaw is determined by astrophysical processes \citep{sampson2015constraining}.
Constraints on the parameters of a given PSD can then inform the underlying physics; for example, an inflationary interpretation can already rule out certain models of inflation \cite{NANOGrav:2023hvm}

Therefore, we may potentially distinguish between GWB origins by comparing spectral shapes and their fit to data.
This is typically done in the context of \emph{Bayesian model comparison}, where the \emph{Bayes factor} (BF) is the metric to compare models.
The BF is the ratio of Bayesian evidences $\mathcal{Z}_\mathcal{M}$ for each model $\mathcal{M}$, where the evidence appears in the denominator of Bayes' theorem,
\begin{equation}
  \label{eq:16}
  p(\theta \mid d, \mathcal{M}) = \frac{p(d \mid\, \theta,\, \mathcal{M})\,p(\theta \mid\, \mathcal{M})}{\mathcal{Z}_\mathcal{M}}, 
\end{equation}
where $\mathcal{Z}_\mathcal{M} = p(d \mid \mathcal{M})$, $d$ is the data, and $\theta$ is the set of model parameters.
For models $\mathcal{M}_1$ and $\mathcal{M}_2$, the BF in favor of $\mathcal{M}_1$ over $\mathcal{M}_2$ is
\begin{equation}
  \label{eq:17}
  \mathrm{BF}_{\mathcal{M}_2}^{\mathcal{M}_1} = \frac{\mathcal{Z}_{\mathcal{M}_1}}{\mathcal{Z}_{\mathcal{M}_2}}.
\end{equation}

PTA analyses typically estimate BFs with one of four techniques \cite{NANOGrav:2023icp}: (1) product-space sampling \cite{carlinBayesianModelChoice1995, arzoumanian2018nanograv,taylor2020bright}, (2) thermodynamic integration \cite{lartillotComputingBayesFactors2006}, (3) likelihood reweighting \cite{hourihaneAccurateCharacterizationStochastic2023}, and (4) computing the Savage-Dickey density ratio \citep{dickeyWeightedLikelihoodRatio1971}. 
Product-space sampling\footnote{Specifically the version of product-space sampling used in PTA analyses, known as ``hypermodeling.'' PTA analyses do not include the ``pseudo-prior'' term used for proposals in the inactive model. \citep{arzoumanian2018nanograv,taylor2020bright}}, likelihood reweighting, and the Savage-Dickey density ratio directly calculates the BF; we cannot recover the evidence $\mathcal{Z}$ from these techniques.
These methods rely on the assumption that the posteriors of the two models under consideration are similar or nested. Estimating the BF becomes increasingly inefficient when the posteriors diverge (for example, see the discussion in Section 3.3 of \citet{Agazie:2026tui}).
Thermodynamic integration, on the other hand, calculates the Bayesian evidence for any model under consideration at the cost of a large increase in computational demand.
There is considerable need for a fast and computationally efficient model comparison method that is applicable to any model of interest, especially given the ever-growing GWB model space \cite{NANOGrav:2023hvm,capriniCosmologicalBackgroundsGravitational2018}.

Nested Sampling \citep[NS,][]{skillingNestedSampling2004,skillingNestedSamplingIdea2024,martinoNestedSamplingCritical2026} is an algorithm designed to calculate the Bayesian evidence directly.
It does so by numerically computing the evidence integral,
\begin{equation}
  \label{eq:14}
  \mathcal{Z} = \int p(d \mid \theta)\, p(\theta)\,\mathrm{d}\theta.
\end{equation}
The rough outline of the algorithm is as follows: First, define a collection of ``live points'' $\hat{\bs{\theta}}$ of size $N_\mathrm{live}$ that spans the prior space.
Next, find the likelihood value for all live points and take the minimum likelihood $L_\mathrm{min}$.
Set the evidence estimate for this $i^\mathrm{th}$ iteration to be
\begin{equation}
  \label{eq:18}
  \mathcal{Z}_i \coloneqq \mathcal{Z}_{i-1} + \left[ \exp\left(\frac{-(i-1)}{N_\mathrm{live}}\right) - \exp\left(\frac{-i}{N_\mathrm{live}}\right) \right] L_{\mathrm{min},\,i},
\end{equation}
with $\mathcal{Z}_0 = 0$, where the bracketed factor is the prior mass enclosed by the shell removed at iteration $i$.
Then, replace the $\hat{\bs{\theta}}_\mathrm{min}$ corresponding to $L_\mathrm{min}$ with a new \(\hat{\bs{\theta}}_\mathrm{new}\) that corresponds to a $L_\mathrm{new} > L_\mathrm{min}$.
Repeat all steps, other than the initial live-point creation, until the estimated evidence remaining in the current live points falls below a user-specified fraction of the accumulated evidence $\mathcal{Z}_i$.
This procedure is equivalent to evaluating the same integral, Eq.~(\ref{eq:14}), as a numerical Lebesgue integral of the un-normalized posterior over enclosed prior mass rather than over the parameters directly.

Unfortunately, this algorithm scales poorly with the number of parameters.
Adequately covering the prior space requires a number of live points that grows super-linearly, and proposing new live points that satisfy the hard likelihood constraint becomes extremely difficult in high dimensions \cite{ashtonNestedSamplingPhysical2022,buchnerNestedSampling2023}.
For PTAs, this means that NS has so far been limited to simple models containing $\sim 10$ pulsars at most \cite{NANOGrav:2023icp} or for noise model selection \citep{chalumeau2022noise}.
To use NS on full arrays of pulsars, we must reduce the dimensionality of the likelihood.
We do this by constructing a PTA likelihood marginalized over all non-GWB processes and training a normalizing flow to act as a fast emulator of this likelihood.

A normalizing flow (NF) is a series of bijective transformations from a simple base distribution to a complex target distribution 
\cite{papamakariosNormalizingFlowsProbabilistic,kobyzevNormalizingFlowsIntroduction2021}.
With an NF, sampling from a target distribution is made trivial by sampling from a base distribution and \emph{pushing forward}.
Calculating probabilities under the target distribution is possible by \emph{pulling back}---i.e., by evaluating their probabilities under the base distribution. 
The goal is to work in the simple base distribution as much as possible by using the NF as a ``translation layer'' to move us to the target.

Given a base distribution $B$, target distribution $T$, and bijective mapping
\begin{equation}
  \label{eq:19}
  f: B \to T,
\end{equation}
the probability of a given $t \in T$ evaluated in the base is
\begin{equation}
  \label{eq:20}
  p_T(t) = p_B\left( f^{-1}(t) \right) \left | \mathrm{det}\left ( J_f \right ) \right | ^{-1},
\end{equation}
where $J_f$ is the Jacobian of $f$.
The map $f$ may be composed of a chain of operations as long as $f$ remains bijective.
We have immense freedom in choosing the operations that constitute $f$, leading to a plethora of potential NF architectures \cite{dinhDensityEstimationUsing2017,durkanNeuralSplineFlows2019}.
In this work, we build $f$ from simple splines and use it to learn the \emph{copula} density of the posterior (see Section~\ref{sec:methods:coppuccino}).
Evaluating $p_T(t)$ with Eq.~(\ref{eq:20}) is cheap, and a flow trained on the posterior can stand in for a likelihood that would otherwise be expensive to evaluate repeatedly.

In this paper, we train a NF on the free-spectrum posterior of a PTA \citep{Lentati:2012xb} marginalized over all non-GWB processes and use it as a fast, low-dimensional surrogate likelihood in the \blackjaxns{} nested sampler.
Our work improves upon the marginalized-likelihood techniques of \citet{lambNeedSpeedRapid2023} by replacing the frequency-independent kernel density estimators with a single flow that captures the correlations between frequencies.
NFs have already proven useful in PTA analyses, both for parameter estimation \cite{vallisneriRapidParameterEstimation2025,Laal:2024trp,Gundersen:2024qmq,gundersen2025escaping} and, recently, for model comparison across GWB sources \cite{Lai:2025xov}.
That model-comparison demonstration was limited to a reduced ten-pulsar array, the same regime NS can already reach.
Because we train our flow on the free-spectrum posterior that is marginalized over all pulsar-specific parameters, the prior-space dimension is set by the number of parameters for the GWB model of interest alone, allowing us to carry evidence-based model comparison to the full array.

\section{Methods}\label{sec:methods}
In this section we describe the computational techniques, statistical tools and derivations, and simulation study design used to develop and validate our proposed model comparison method.
This section is organized as follows:
\begin{itemize}
    \item Section \ref{sec:methods:coppuccino} introduces \coppuccino{}, the NF software package we use throughout this study.
    \item Section \ref{sec:methods:flow-accel} describes the process of creating a NF-based likelihood, handling the tails of the distribution, mathematical guarantees that the Bayes factors are unbiased as compared to traditional likelihoods, and how the flow is integrated into a nested sampling routine.
    \item Section \ref{sec:methods:validation} validates our approach on a target where the Bayesian evidence is known in closed form.
    \item Section \ref{sec:methods:sims} outlines the simulation study used to validate our method on realistic PTA data.
\end{itemize}

\subsection{\coppuccino{}}\label{sec:methods:coppuccino}
\coppuccino is a \textsc{Python} package that uses NFs to estimate a probability density $p(\mathbf{x})$ with $\mathbf{x} = (x_1, x_2, \ldots, x_d)\in \mathbb{R}^d$ given samples from the density function.
Once fitted, the model can draw new samples and report the probability density at a given point.
\coppuccino is built on top of \flowjax~\cite{ward2023flowjax}, a package for producing NFs using \jax.

We write the marginal cumulative distribution function (CDF) of coordinate $i$ as $F_i$ and its density as $p_i = F_i'$.
The marginal CDF maps each coordinate $x_i$ to $u_i = F_i(x_i)\sim U[0, 1]$.
After this transformation, all marginals are uniformly distributed, but the dependence structure between these uniform distributions persists. The joint CDF of the uniform marginals is called a \textit{copula} $C$ and has density $c$ which encodes the dependence structure between the marginals.
Sklar's theorem~\citep{sklarFonctionsRepartitionDimensions1959} states that the joint CDF $H(\mathbf{x})$ can be written as
\begin{equation}
    H(\mathbf{x}) = C(F_1(x_1),\ldots,F_d(x_d)),
\end{equation}
and taking a derivative leads to a product of marginals and the dependence structure
\begin{equation}
    p(\mathbf{x}) = c(F_1(x_1),\ldots,F_d(x_d))\prod_{i=1}^d p_i(x_i).
\end{equation}

Instead of fitting the joint distribution with a NF directly, we transform the marginals using their empirical CDF $\hat{F}_i(x_i)$ and fit the copula density $c$.
The empirical CDF is computed from the samples and fit with a monotonic rational-quadratic spline.
We then use the empirical CDF as a map from the parameter space to the unit cube $[0, 1]^d$.
While the unit cube ideally yields a tractable copula density, we find that the number of parameters in the flow is reduced by transforming to a standard normal.
We use the inverse Gaussian CDF to map the uniformly distributed parameters to a standard normal distribution $z_i = \Phi^{-1}(u_i)$.
The samples then live on $\mathbb{R}^{d}$, leaving the dependence structure unbounded.
The copula density rarely deviates from Gaussian in problems with weak correlation structure, allowing a small number of parameters to effectively fit it.
Working in this parameterization also removes multimodality from the marginals.

Fitting proceeds in two stages.
First, the marginals are estimated once with a one-dimensional empirical CDF built directly from each parameter's sorted samples using monotonic rational-quadratic splines and then held fixed.
$n$ knots are placed at equal intervals spanning quantiles $0$ to $1$, with $n = \max\left(20,\, \min(n_{\max},\, \lfloor N/3 \rfloor)\right)$, where $N$
is the number of data points and $n_{\max} = 200$ by default.
Within \coppuccino, we create a fit to the tails of the distribution; however, in this work, we instead truncate the quantiles at 2\% and 98\% and fit tails as described in Sec.~\ref{sec:tails}.

Next, we fit a NF to the copula density, the dependence structure between the Gaussianized marginals, using \flowjax.
We learn the flow, an invertible transformation $T$ with adjustable parameters, from a $d$-dimensional standard normal base distribution $\mathcal{N}(\mathbf{0}, I_d)$ to the Gaussianized posterior $\mathbf{z}$.
$T$ is produced through a trainable triangular spline flow architecture.
Since $T$ is invertible with a tractable Jacobian, it defines an exact density on $\mathbf{z}$ through the change of variables,
\begin{equation}
    \log p_Z(\mathbf{z})
    = \log \mathcal{N}\!\left(T^{-1}(\mathbf{z})\mid\, \mathbf{0}, I_d\right)
    + \log\left|\det \frac{\partial T^{-1}}{\partial \mathbf{z}}\right|,
    \label{eq:flow_density}
\end{equation}
where the second term accounts for the change in volume induced by the transformation.

The triangular spline flow is a composition of $L$ identical layers ($L = 6$ by default).
First, every coordinate is passed through its own monotonic rational-quadratic spline with $K$ knots ($K = 4$ by default).
These splines are defined on $[-1, 1]$, so each coordinate is mapped into the interval using a leaky $\tanh$ on $[-3, 3]$ with linear tails.
The spline is applied and the mapping is inverted so the layer output again spans the real line.
Coordinates falling outside the spline interval are passed through unchanged, leaving the extreme tails undistorted.
Because this step acts on each coordinate separately, the Jacobian is diagonal.
Next, the coordinates are mixed by a lower-triangular affine map $\mathbf{z} \mapsto \mathbf{A}\mathbf{z} + \mathbf{b}$.
The diagonal entries of this map are constrained to be positive and the rows are weight-normalized for training stability.
Since $\mathbf{A}$ is triangular, $\log|\det \mathbf{A}| = \sum_i \log A_{ii}$, so each layer's log-determinant is a sum of $d$ scalar terms rather than a $d \times d$ determinant.
Finally, there is an untrained permutation that reorders the coordinates.
Since a lower-triangular map cannot allow the first coordinate to depend on others, reordering allows arbitrary parameter dependence over multiple layers.
With these default settings, the flow contains $2548$ trainable parameters for $d = 14$.

Composing the two stages gives the density of the fitted model in closed form.
The marginal map ${z_i = \Phi^{-1}(\hat{F}_i(x_i))}$ acts on each coordinate
separately, so its Jacobian is diagonal and contributes a sum of one-dimensional terms,
\begin{equation}
    \log \hat{p}(\mathbf{x}) = \log p_Z(\mathbf{z})
    + \sum_{i=1}^{d}\left[\log \hat{p}_i(x_i) - \log \varphi(z_i)\right],
    \label{eq:coppuccino_density}
\end{equation}
where $\hat{p}_i = \hat{F}_i'$ is the estimated marginal density and $\varphi$ is the standard normal density.
The first term is given by the flow through Eq.~\eqref{eq:flow_density}, and the sum is the Jacobian of the Gaussianization.
Because the marginal transforms are held fixed, this sum is independent of the flow parameters, so maximizing the likelihood of the model is equivalent to maximizing the likelihood of the flow on the Gaussianized samples.

The flow is trained on the Gaussianized samples by minimizing the negative log-likelihood of the transformed data.
We use an Adam optimizer \citep{kingma2014adam} with a learning rate of $1\times10^{-3}$.
Training is performed on 90\% of the data with 10\% held out as a validation set.
We train for a maximum of 400 epochs where each epoch consists of one complete pass over the training data, where the data are used in mini batches of 100 samples at a time.
If after 30 consecutive epochs of training, the validation loss has not improved, the training stops and the parameters from the best epoch, by validation loss, are used.

Put simply, \coppuccino splits the density estimation into two pieces that are easier to solve separately than together.
The marginals are estimated directly from the sorted samples so that part of the fit requires no training at all.
The flow is left with only the dependence structure to learn.
A flow fit to the joint density must capture the same dependence, but it must also learn the marginals too, which can require many more parameters when the marginal shapes are complicated.

\subsection{Flow-accelerated nested sampling}\label{sec:methods:flow-accel}
The main purpose of our method is to construct a low-dimensional, computationally-inexpensive PTA likelihood that is ideal for NS.
This is in contrast to traditional PTA likelihoods that are high-dimensional and computationally expensive.
We accomplish this by training an NF on posterior samples of a PTA free-spectrum likelihood (marginalized over non-GWB parameters) and using the flow as a fast likelihood emulator in a nested sampler.

\subsubsection{PTA free-spectrum analysis}
A PTA analysis models the timing residuals $\bs{\delta t}$ as a zero-mean Gaussian process, so the likelihood is a multivariate Gaussian (Eq.~\ref{eq:11}) whose covariance is the sum of a deterministic timing model, per-pulsar white and red noise, and a common process shared across pulsars that carries the GWB.
The red-noise and GWB contributions are represented in a Fourier basis at the frequencies $f_i$ defined below, so the GWB enters the likelihood only through the power it contributes at each $f_i$.
We refer the reader to \citet{NANOGrav:2023icp} for a complete description of the PTA likelihood and its implementation.

Starting with a PTA data set, we first run a free-spectrum analysis \citep{Lentati:2012xb}, which models the GWB power at each $f_i$ without assuming a spectral shape.
We define $N_f$ frequencies with the $i^{\textrm{th}}$-frequency given by $ f_i = i / T_\mathrm{total}$, where $T_\mathrm{total}$ is the total time-span of the PTA dataset.
The PSD of the free-spectrum process is then
\begin{equation}
  \label{eq:4}
  S(f_i, \bs{\rho}) = \frac{\delta_{ij} \rho_j^2}{\Delta f_j}.
\end{equation}
In words, the PSD evaluated at the $i^{\textrm{th}}$-frequency is proportional to the $i^{\textrm{th}}$-component of the amplitude vector $\bs{\rho}$, squared.

Under the assumption that the GWB is a zero-mean GP with variance $\bs{\rho}^2$, the free-spectrum analysis then gives us a posterior on that variance.
This variance is the expected power in the GWB over an ensemble of universes.

The free-spectrum is useful because it is agnostic to any specific spectral shape, i.e.,\, it is agnostic to the source(s) of the GWB.
As long as the source is also a GP, the free spectrum contains it as a nested sub-model.
We can then use the free-spectrum posteriors on the PSD parameters \(\bs{\rho}\) as a marginalized and compressed version of the PTA likelihood, an approach pioneered in \citet{lambNeedSpeedRapid2023}.

\subsubsection{The surrogate likelihood}
To construct a likelihood from the free-spectrum posteriors, we first write down the relationship between the posterior and the likelihood using Bayes' theorem,
\begin{equation}
  \label{eq:freespec-posterior}
  p(\bs{\rho} \mid \bs{\delta t}) = \frac{p(\bs{\delta t} \mid \bs{\rho})\, p(\bs{\rho})}{\mathcal{Z}_\mathrm{fs}}, 
\end{equation}
where $\bs{\delta t}$ contains the observed timing residuals.
Up to a normalizing constant, the posterior is equal to the likelihood times the prior.

The likelihood $p(\bs{\delta t} \mid \bs{\rho})$ here is already marginalized over every non-GWB parameter.
The free-spectrum MCMC samples the joint posterior over the amplitudes $\bs{\rho}$ and all other parameters $\bs{\eta}$, and we keep the marginal in $\bs{\rho}$:
\begin{align}
  \label{eq:posterior-marg}
  p(\bs{\rho} \mid \bs{\delta t}) &= \int p(\bs{\rho}, \bs{\eta} \mid \bs{\delta t})\, \mathrm{d}\bs{\eta},\\
                                  &= \frac{p(\bs{\rho})}{\mathcal{Z}_{\mathrm{fs}}} \int p(\bs{\delta t} \mid \bs{\rho}, \bs{\eta})\, p(\bs{\eta})\,\mathrm{d}\bs{\eta},\\
  \label{eq:likelihood-marg}
  &\propto p(\bs{\delta t} \mid \bs{\rho} ),
\end{align}
where $\mathcal{Z}_\mathrm{fs}$ is the evidence of the free-spectrum model, and we drop the prior term $p(\bs{\rho})$ in Eq.~(\ref{eq:likelihood-marg}) because it is a constant, Uniform prior\footnote{This method does not require Uniform priors. If non-Uniform priors on $\bs{\rho}$ are used, then we must carry around the $p(\bs{\rho})$ term, a trivial extension of the method.}.
We sample $\bs{\eta}$ from its posterior, but marginalizing the joint posterior over $\bs{\eta}$ is the same as integrating the likelihood against the prior $p(\bs{\eta})$, as seen in Eq.~(\ref{eq:posterior-marg}).
Eq.~(\ref{eq:likelihood-marg}) gives us the \emph{marginalized PTA likelihood} that depends only on $\bs{\rho}$, and it is directly proportional to the posterior $p(\bs{\rho} \mid \bs{\delta t})$.

We can therefore use the posterior density as a surrogate for the marginalized likelihood.
To do so, we need a way to evaluate the posterior probability of a given $\bs{\rho}$.
We train a normalizing flow on the posterior $p(\bs{\rho} \mid \bs{\delta t})$ using \coppuccino{}, which gives us a fast, accurate way to evaluate posterior probabilities and to draw posterior samples.

\subsubsection{Taming the tails}
\label{sec:tails}
The flow is an accurate surrogate likelihood inside the support of its training samples.
However, the nested sampler will routinely query it outside the training data because the live points span the entire prior, which is often wider than the posterior.
Extending our likelihood surrogate into these regions can be tricky and requires close attention.
Previous methods \citep{lambNeedSpeedRapid2023} appended a constant likelihood beyond the support of the training data---holding the likelihood at its boundary value.
Here, we improve upon this method by exploiting our knowledge of the asymptotic behavior of the PTA free-spectrum likelihood as $\bs{\rho} \to 0$ and $\bs{\rho} \to \infty$.

At high power, the PTA likelihood decays exponentially.
Conversely, at low power, it plateaus at the noise floor.
We can read this off directly from the Gaussian form of the PTA likelihood:
\begin{equation}
  \label{eq:11}
  p(\bs{\delta t} \mid \bs{\theta}) = \mathcal{N} \left( \bs{\delta t} \mid 0,\, \bs{C}\left(\bs{\theta}\right) \right),
\end{equation}
where $\bs{\delta t}$ are the timing residuals and $\bs{C}\left(\bs{\theta}\right)$ is the covariance matrix as a function of the free parameters.
The GWB parameters enter Eq.~(\ref{eq:11}) only through the covariance, while the observed residuals stay fixed.
Therefore, both limits follow from how $\bs{C}$ depends on $\bs{\rho}$.

We adopt the notation of \citet{NANOGrav:2023icp} and write $\bs{C} = \bs{N} + \bs{T} \bs{B} \bs{T}^{T}$.
Here $\bs{N}$ is the white-noise covariance, and $\bs{T} = \left[\bs{M}\; \bs{F}\right]$ collects the timing-model design matrix and the Fourier basis.
The prior covariances for the timing model and the Fourier-coefficient covariance are contained in the block-diagonal matrix $\bs{B}$.
We assume an improper prior on the timing model covariance such that its entries go to zero in $\bs{B}^{-1}$.
Therefore, we are only interested in the non-timing-model portions of $\bs{B}$, with block diagonal entries given by
\begin{equation}
  \label{eq:phi-blocks}
  \left[\bs{\phi}\right]_{(ai)(bj)} = \delta_{ij} \left( \delta_{ab}\, \varphi_{ai} + \Gamma_{ab} \Phi_i \right),
\end{equation}
where the intrinsic red noise of pulsar $a$ at frequency $i$ is $\varphi_{ai}$, the overlap reduction function $\Gamma_{ab}$ equals $\delta_{ab}$ for the common, uncorrelated red-noise (CURN) process we fit, and the common power in bin $i$ is $\Phi_i = S(f_i) \Delta f = \rho_i^2$.

Two pieces of the log-likelihood depend on $\rho_i$: the quadratic form $-\tfrac{1}{2}\bs{\delta t}^{T} \bs{C}^{-1} \bs{\delta t}$ and the log-determinant $-\tfrac{1}{2}\log\det \bs{C}$.
For the determinant, the matrix determinant lemma gives
\begin{equation}
  \label{eq:logdet-split}
  \log\det \bs{C} = \log\det \bs{N} + \log\det \bs{B} - \log\det \bs{\Sigma},
\end{equation}
where $\bs{\Sigma} = \left( \bs{B}^{-1} + \bs{T}^{T} \bs{N}^{-1} \bs{T} \right)^{-1}$.
As $\rho_i \to \infty$, $\log\det \bs{B}$ grows as $4 N_\mathrm{psr} \log \rho_i$, where the extra factor of two comes from the sine and cosine components both modeled by $\rho_i$.

In the same high-power limit, $\bs{\Sigma}$ and the quadratic form become independent of $\rho_i$, so $\log\det \bs{B}$ is the only term that still grows.
Because the log-determinant enters the log-likelihood as $-\tfrac{1}{2}\log\det \bs{C}$, and assuming a vanishing covariance with this high-power mode, the logarithmic derivative of the log-likelihood with respect to this mode is then
\begin{equation}
  \label{eq:upper-slope}
  \frac{\partial \log p(\bs{\delta t} \mid \rho_i)}{\partial \log \rho_i} = -2 N_\mathrm{psr},
\end{equation}
and this holds for all non-singular $\bs{\Gamma}$ \footnote{In general the slope is $-2\,\mathrm{rank}(\bs{\Gamma})$ rather than $-2 N_\mathrm{psr}$.
  The $\rho_i$-dependent block of $\bs{B}$ is $\bs{D} + \Phi_i \bs{\Gamma}$, with $\bs{D} = \mathrm{diag}(\varphi_{ai})$ the intrinsic red noise.
  We can whiten the correlation matrix as $\bs{D}^{-1/2} \bs{\Gamma} \bs{D}^{-1/2}$, whose eigenvalues we call $\mu_k$.
  Then, $\bs{B}$ is $\bs{D^{1/2}} (\bs{I} + \Phi_i \bs{D}^{-1/2} \bs{\Gamma} \bs{D}^{-1/2})\bs{D^{1/2}}$.
  Taking the determinant and using ${\det(\bs{A}\bs{B}) = \det \bs{A} \det \bs{B}}$ gives $\det \bs{D} \prod_k \left( 1 + \Phi_i \mu_k \right)$ for $\det\bs{B}$.
  Differentiating with respect to $\log \rho_i$ therefore picks up only the $\mathrm{rank}(\bs{\Gamma})$ terms with $\mu_k \neq 0$, each approaching $2$ as $\Phi_i \to \infty$ and picking up an extra factor of 2 for the sine and cosine components.
  Both $\bs{\Gamma} = \bs{I}$ and the Hellings-Downs matrix are full rank for distinct sky positions, so the slope is the same in both cases.}.
In the opposite limit as $\rho_i \to 0$, $\Phi_i$ falls below $\varphi_{ai}$, and $\bs{\phi}$ then depends only on the intrinsic red noise.
Neither $\bs{C}$ nor the likelihood changes further, so the lower tail plateaus.

The upper tail approaches Eq.~(\ref{eq:upper-slope}) only asymptotically, and near the edge of the training data the decay is much shallower.
Rather than impose the asymptotic slope directly at the edge of the training data, we freeze all nuisance parameters at their posterior medians and explore the behavior of each bin's upper tail with the full likelihood.
Because the covariance is a sum of non-negative variances, increased common power cannot be entirely canceled by the per-pulsar noise.
This justifies holding the nuisance parameters constant, as they do not significantly affect the likelihood in this high-power regime.

\begin{figure}
  \includegraphics[width=\columnwidth]{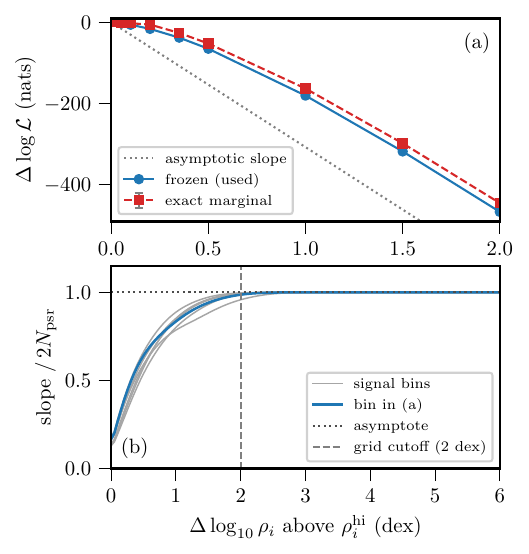}
  \caption{
    Upper-tail behavior above the trusted-region edge $\rho_i^\mathrm{hi}$, for a power-law realization.
    \textbf{(a)} Profile for one signal-dominated bin.
    Freezing the nuisance parameters at their posterior medians (solid) tracks the exact marginal profile from NUTS and Fisher's identity (dashed).
    The asymptotic slope of Eq.~(\ref{eq:upper-slope}) (dotted) is far steeper than the profile here, which is why we measure the profile rather than impose the asymptote from the boundary.
    \textbf{(b)} Local slope of the frozen profile in units of the predicted asymptote, for every signal-dominated bin, on a grid extending well beyond the range of panel (a).
    The magnitude of the slope in each bin converges to $2 N_\mathrm{psr}$.
  }
  \label{fig:upper-tail}
\end{figure}

We evaluate the likelihood on a grid of $\rho_i$'s out to two dex (path sampling, \citep{gelmanSimulatingNormalizingConstants1998}), interpolate during excursions above the training data, and continue along Eq.~(\ref{eq:upper-slope}) beyond the training data.
Fig.~\ref{fig:upper-tail}a validates this prescription against the exact marginal found by NUTS \citep{hoffmanNoUTurnSamplerAdaptively} sampling the full likelihood while holding $\rho_i$ constant at each grid point.
We choose to use this frozen upper tail instead of the exact marginal for computational savings in our large-scale study.
However, the full marginal is preferable for production-level results.

For the lower tails, our treatment depends on whether the bin is noise-dominated.
If it is, the sampler will explore the noise floor, and the surrogate is adequate.
Signal-dominated bins, however, will never sample the noise floor.
The decrease in probability from the edge of the training data to the noise floor is the penalty a model pays for predicting power well below the level the data support, and it must be properly characterized for reliable model comparison.
Unlike the upper tails, we cannot hold the nuisance parameters at their posterior medians.
In the lower tails, the nuisance parameters can ``inflate'' to make up for low common-process power, so they are strongly covariant with $\bs{\rho}$.

For each of the signal-dominated bins, we again define a grid of $\bs{\rho}$'s extending two dex below the training data and marginalize over the nuisance parameters numerically.
We utilize Fisher's identity (see Eq. (5.12) of \citep{cappeMaximumLikelihoodInference2005}) to relate the gradient of the marginalized log-likelihood to the expected gradient of the full log-likelihood over the posterior of the nuisance parameters $\bs{\eta}$:
\begin{equation}
  \label{eq:fisher-identity}
  \frac{\mathrm{d} \log p(\bs{\delta t} \mid \rho_i)}{\mathrm{d} \rho_i} = \left\langle \frac{\partial \log p(\bs{\delta t} \mid \rho_i,\, \bs{\eta})}{\partial \rho_i} \right\rangle_{p(\bs{\eta} \mid \rho_i,\, \bs{\delta t})}.
\end{equation}

To acquire posterior samples, we hold the signal-dominated bin's $\rho_i$ constant on the grid and allow all other parameters to vary during a NUTS run, and evaluate the gradients cheaply by automatic differentiation.
Integrating Eq. (\ref{eq:fisher-identity}) gives the change in likelihood between grid points, which we interpolate between.
Below the grid for signal-dominated bins and the training data of the noise-dominated bins, we hold the likelihood value constant, having assured ourselves that the grid or the training data extend to low enough power such that the likelihood has reached the noise floor.

In order to avoid edge-effects, the boundaries that we splice our tails onto, $[\rho_i^\mathrm{lo}, \rho_i^\mathrm{hi}]$, are the $2\%$ and $98\%$ quantiles of the training data.
Outside these bounds, we clip $\rho_i$ to the nearest boundary, evaluate the flow there, and add the change in log-likelihood measured along that bin's profile between the edge and the requested value.
Define the following operation to restrict a given $\bs{\rho}$ within the training data's $[\rho_i^\mathrm{lo}, \rho_i^\mathrm{hi}]$,
\begin{equation}
  \label{eq:15}
  \mathrm{clip}(\bs{\rho}) = \mathrm{min}\left(\bs{\rho}^\mathrm{hi},\, \mathrm{max}\left(\bs{\rho}^\mathrm{lo},\,\bs{\rho}\right)\right).
\end{equation}
The final surrogate log-likelihood is then
\begin{equation}
\begin{split}
  \label{eq:tail-splice}
  \log p\left(\bs{\delta t} \mid \bs{\rho}\right) &=  \log q\left(\mathrm{clip}(\bs{\rho})\right) \\
  &+ \sum_{i} \left[\log p(\bs{\delta t} \mid \rho_i) - \log p(\bs{\delta t} \mid \mathrm{clip}(\bs{\rho}))\right],
\end{split}
\end{equation}
where $q$ is the flow density and the clip acts component-wise.
Each difference vanishes for a bin inside the defined boundaries, so the tails inherit the flow's own constant offset rather than introducing one of their own.

Summing the per-bin corrections in Eq.~(\ref{eq:tail-splice}) treats simultaneous excursions into the tails as independent.
For the upper tail, this is treatment is well justified because the common power dominates the variance, leading to Eq.~(\ref{eq:upper-slope}).
For the lower tail, it is an approximation because the nuisance parameters that can absorb the missing power often influence multiple bins at once.
Thus, we verify the full scheme numerically in Section~\ref{sec:results:model-selection}. 

\subsubsection{Sampling the models}
Now that we have a likelihood surrogate that is well-behaved everywhere, we can supply it to a NS code.
We use \blackjaxns{}: a recently developed algorithm from~\citet{Yallup:2025sty} based on the \textsc{blackjax} Bayesian-inference library~\citep{cabezas2024blackjax}.
This variant of NS is written in \jax{} \cite{jax2018github}, which enables easy GPU acceleration, automatic vectorization, and interoperability with \jax{}'s growing scientific-computing ecosystem.
\blackjaxns{} uses \jax{} to update its live points \emph{in parallel}, giving a large speedup compared to other NS codes and making especially effective use of GPU acceleration.
This sampler gives us posterior constraints and evidences in minutes on consumer hardware.

To test new models with the likelihood surrogate and \blackjaxns{}, we construct a hierarchical model on the free-spectrum parameters $\bs{\rho}$.
First, we choose model(s) $\mathcal{M}$ with parameters $\bs{\theta}$.
Then, the hierarchical model is
\begin{equation}
  \label{eq:7}
  p(\bs{\rho}, \bs{\theta} \mid \bs{\delta t}, \mathcal{M}) \propto  p(\bs{\delta t} \mid \bs{\rho})\, p(\bs{\rho} \mid \mathcal{M}, \bs{\theta})\, p(\bs{\theta} \mid \mathcal{M})\,,
\end{equation}
where $p(\bs{\delta t} \mid \bs{\rho})$ is our NF likelihood surrogate.

Note that $p(\bs{\rho} \mid \mathcal{M}, \bs{\theta})$ is a Dirac-delta function under the Gaussian GWB assumption: it is zero everywhere except at $\bs{\rho} = \mathcal{M}(\bs{\theta})$, where $\mathcal{M}(\bs{\theta})$ denotes the evaluation of model $\mathcal{M}$'s spectrum with parameters $\bs{\theta}$.
The model $\mathcal{M}$ may be a single source or a \emph{compound model} whose PSD is the sum of several component spectra, e.g. a cosmological background in conjunction with an astrophysical background from supermassive black hole binaries.
In that case $\bs{\theta}$ collects the parameters of every component and $\mathcal{M}(\bs\theta)$ returns their summed power.
The delta-function relation $\bs\rho = \mathcal{M}(\bs\theta)$ is unchanged, so compound models require no special treatment.

Therefore, we never need to \emph{sample} $\bs{\rho}$\,.
We instead equivalently \emph{calculate} $\bs{\rho} = \mathcal{M}(\bs{\theta})$ and evaluate the probability for that $\bs{\rho}$ with the likelihood surrogate.
Then, rewriting Eq.~(\ref{eq:7}) with this notation and simplifying leads us to
\begin{equation}
  \label{eq:13}
  p(\bs{\rho}, \bs{\theta} \mid \bs{\delta t},\, \mathcal{M}) \propto  p(\bs{\delta t} \mid \bs{\rho})\,
  \delta(\bs{\rho}-\mathcal{M}(\bs{\theta}))\, p(\bs{\theta} \mid \mathcal{M}).
\end{equation}

To compare two models $\mathcal{M}_i$ and $\mathcal{M}_j$, we run the nested sampler for each and gather their evidences.
The Bayes factor is then just the ratio of evidences or the difference of their logarithms,
\begin{equation}
  \label{eq:9}
  \logten \mathrm{BF}^{\mathcal{M}_i}_{\mathcal{M}_j} = \logten{\mathcal{Z}_{\mathcal{M}_i}} - \logten{\mathcal{Z}_{\mathcal{M}_j}}.
\end{equation}

\subsubsection{Marginalization leaves the Bayes factor invariant}\label{sec:methods:marginalization}
Using this marginalized likelihood in place of the full one does not change the Bayes factor.
To see this, we will build a hierarchical model as in the previous section on $\bs{\rho}$ with parameters $\bs{\theta}$ and use a typical PTA free-spectrum likelihood.
We partition the parameters of a model $\mathcal{M}$ into the GWB-related parameters $\bs{\theta}$, the free-spectrum parameters $\bs{\rho}$, and all other parameters $\bs{\eta}$.

The evidence for a model $\mathcal{M}$ is
\begin{equation}
  \label{eq:full-evidence}
  \mathcal{Z}_\mathcal{M} = \iiint p(\bs{\delta t} \mid \, \bs{\rho}, \bs{\theta}, \bs{\eta}, \mathcal{M} )\, p(\bs{\rho}, \bs{\theta}, \bs{\eta} \mid \mathcal{M})\, \mathrm{d} \bs{\rho}\, \mathrm{d} \bs{\theta}\, \mathrm{d} \bs{\eta}\,.
\end{equation}
In all realistic PTA models, there is no conditional probability between $\bs{\rho}$ and $\bs{\eta}$ or $\bs{\theta}$ and $\bs{\eta}$, but there is conditional dependence of $\bs{\rho}$ and $\bs{\theta}$.
The parameters $\bs{\theta}$ together with the PSD of model $\mathcal{M}$ \emph{predict} a set of free-spectrum parameters $\bs{\rho}$: this is the hierarchical part of our model.
Therefore,
\begin{equation}
  \label{eq:21}
  p(\bs{\rho},\, \bs{\theta},\, \bs{\eta} \mid\, \mathcal{M}) = p(\bs{\theta} \mid\, \bs{\rho},\, \mathcal{M})\, p(\bs{\rho})\, p(\bs{\eta})\,,
\end{equation}
where the free-spectrum prior $p(\bs{\rho})$ and the nuisance prior $p(\bs{\eta})$ carry no dependence on $\mathcal{M}$.
Eq.~(\ref{eq:full-evidence}) then becomes,
\begin{widetext}
\begin{equation}
  \label{eq:6}
  \mathcal{Z}_\mathcal{M} = \iiint p(\bs{\delta t} \mid \, \bs{\rho},\, \bs{\theta},\, \bs{\eta},\, \mathcal{M} )\, p(\bs{\theta} \mid \bs{\rho},\, \mathcal{M})\, p(\bs{\rho})\, p(\bs{\eta})\, \mathrm{d} \bs{\rho}\, \mathrm{d} \bs{\theta}\, \mathrm{d} \bs{\eta}\,.
\end{equation}
\end{widetext}
A PTA free-spectrum likelihood, however, does not depend on the parameters $\bs{\theta}$ or the choice of model $\mathcal{M}$---it only depends on $\bs{\rho}$ (and the nuisance parameters $\bs{\eta}$).
Thus, we may drop $\bs{\theta}$ and $\mathcal{M}$ from the likelihood term as there is no conditional dependence:
\begin{equation}
  \label{eq:23}
  p(\bs{\delta t} \mid \, \bs{\rho},\, \bs{\theta},\, \bs{\eta},\, \mathcal{M} ) = p(\bs{\delta t} \mid \, \bs{\rho},\, \bs{\eta} ).
\end{equation}

Then, using Eq.~(\ref{eq:23}) together with Bayes' theorem, we may rewrite Eq.~(\ref{eq:6}) as
\begin{equation}
  \label{eq:22}
  \mathcal{Z}_\mathcal{M} = \mathcal{Z}_\mathrm{fs} \iiint p(\bs{\rho}, \bs{\eta} \mid \bs{\delta t})\, p(\bs{\theta} \mid \bs{\rho}, \mathcal{M})\, \mathrm{d} \bs{\rho}\, \mathrm{d} \bs{\theta}\, \mathrm{d} \bs{\eta}\,.
\end{equation}
Integrating over $\eta$ leaves us with,
\begin{equation}
  \label{eq:24}
  \mathcal{Z}_\mathcal{M} = \mathcal{Z}_\mathrm{fs} \iint p(\bs{\rho} \mid\, \bs{\delta t})\, p(\bs{\theta} \mid\, \bs{\rho},\, \mathcal{M})\, \mathrm{d} \bs{\rho}\, \mathrm{d} \bs{\theta}\,,
\end{equation}
where $p(\bs{\rho} \mid \bs{\delta t})$ is the density our \coppuccino{} flows learn.

To integrate over $\bs{\rho}$, we expand the conditional $p(\bs{\theta} \mid\, \bs{\rho},\, \mathcal{M})$ with Bayes' theorem.
The forward map is deterministic: model $\mathcal{M}$ with parameters $\bs{\theta}$ predicts a single free spectrum $\bs{\rho} = \mathcal{M}(\bs{\theta})$, where $\mathcal{M}(\bs{\theta})$ denotes the evaluation of model $\mathcal{M}$'s spectrum at parameters $\bs{\theta}$.
Thus, $p(\bs{\rho} \mid \bs{\theta},\, \mathcal{M})$ is a Dirac delta and
\begin{align}
  \label{eq:25}
  p(\bs{\theta} \mid \bs{\rho}, \mathcal{M}) &= \frac{p(\bs{\rho} \mid \bs{\theta}, \mathcal{M})\, p(\bs{\theta} \mid \mathcal{M})}{p(\bs{\rho})},\\
                                  &= \frac{\delta\left(\bs{\rho} - \mathcal{M}(\bs{\theta})\right)\, p(\bs{\theta} \mid \mathcal{M})}{p(\bs{\rho})}.
\end{align}
Substituting into Eq.~(\ref{eq:24}) and integrating the delta over $\bs{\rho}$ gives
\begin{align}
  \label{eq:26}
  \mathcal{Z}_\mathcal{M} &= \mathcal{Z}_\mathrm{fs} \int \frac{p(\mathcal{M}(\bs{\theta}) \mid\, \bs{\delta t})\, p(\bs{\theta} \mid\, \mathcal{M})}{p(\mathcal{M}(\bs{\theta}))}\,  \mathrm{d} \bs{\theta}\,,\\
 & = \frac{\mathcal{Z}_\mathrm{fs}}{c} \int p(\mathcal{M}(\bs{\theta}) \mid\, \bs{\delta t})\, p(\bs{\theta} \mid \mathcal{M})\,  \mathrm{d} \bs{\theta}\,,
\end{align}
where the Uniform free-spectrum prior $p(\bs{\rho})=c$.

When comparing models using the Bayes factor, the constant factor $\mathcal{Z}_\mathrm{fs}/c$ cancels\footnote{If instead $\mathcal{M}_1$ and $\mathcal{M}_2$ correspond to different overlap reduction functions (ORFs) $\Gamma_1$ and $\Gamma_2$, the constant factor no longer reduces to the identity.
The constant prior term will still cancel, but we must multiply Eq.~(\ref{eq:ns-target}) by a pre-factor of ${\mathcal{Z}_{\mathrm{fs},\,\Gamma_1}}/{\mathcal{Z}_{\mathrm{fs},\,\Gamma_2}}$.
This is simply the Bayes factor for ORF $\Gamma_1$ over $\Gamma_2$ using a free-spectrum model.
Additionally, models assuming a specific ORF must use a likelihood surrogate produced assuming the same ORF.}:
\begin{align}
  \label{eq:27}
  \mathrm{BF}^{\mathcal{M}_1}_{\mathcal{M}_2}&= \frac{\mathcal{Z}_{\mathcal{M}_1}}{\mathcal{Z}_{\mathcal{M}_2}}, \\
&= \frac{\mathcal{Z}_\mathrm{fs}/c \int p(\mathcal{M}_1(\bs{\theta}) \mid\, \bs{\delta t})\, p(\bs{\theta} \mid\, \mathcal{M}_1)\,  \mathrm{d} \bs{\theta}}{\mathcal{Z}_\mathrm{fs}/c \int p(\mathcal{M}_2(\bs{\theta}) \mid\, \bs{\delta t})\, p(\bs{\theta} \mid\, \mathcal{M}_2)\,  \mathrm{d} \bs{\theta}}, \\
\label{eq:ns-target}
&= \frac{\int p(\mathcal{M}_1(\bs{\theta}) \mid\, \bs{\delta t})\, p(\bs{\theta} \mid\, \mathcal{M}_1)\,  \mathrm{d} \bs{\theta}}{\int p(\mathcal{M}_2(\bs{\theta}) \mid\, \bs{\delta t})\, p(\bs{\theta} \mid\, \mathcal{M}_2)\,  \mathrm{d} \bs{\theta}}\,.
\end{align}
For each model $\mathcal{M}$, Eq.~(\ref{eq:ns-target}) is the \emph{exact} integral we compute with our method.
Therefore, our method is analytically equivalent to the full PTA evidence, and the only differences must come from the flow fidelity, inexactness of the nested sampling algorithm, or treatment of the tails.

\subsection{Validation on an analytic target}\label{sec:methods:validation}
To validate our method, we apply it to a target whose evidence we know analytically: a ten-dimensional Gaussian.
We use a standard multivariate Gaussian, so the likelihood is simply the standard-normal probability density function (PDF).
NS requires that we specify a prior range for the parameters of our likelihood, and we choose a uniform prior over the hypercube bounded by $-5$ to $5$ in each dimension.

This choice makes the evidence integral a simple difference of the standard-normal CDF evaluated at the bounds, and the likelihood factorizes over the dimensions because the covariance is diagonal,
\begin{equation}
  \label{eq:gaussian-evidence}
  \mathcal{Z} = \int\limits_{[-5,\,5]^{D}} \prod_{i=1}^{D} \mathcal{N}(x_i \mid\, 0,\, 1)\, p(x_i)\, \mathrm{d}x_i
  = \left[ \frac{\Phi(5) - \Phi(-5)}{10} \right]^{D},
\end{equation}
where $D = 10$, $\Phi$ is the standard normal CDF, and $p(x_i) = 1/10$ is the uniform prior density on $[-5, 5]$.

\begin{figure}
  \includegraphics[width=\columnwidth]{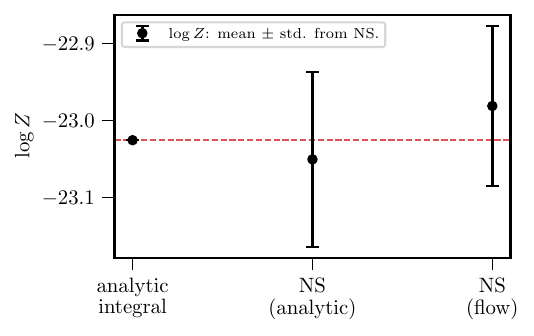}
  \caption{
    Evidence validation on a ten-dimensional Gaussian target.
    We sample a standard multivariate Gaussian with a uniform hypercube prior on the ten-dimensional random vector $x_i \sim \mathcal{U} (-5 , 5) \; \mathrm{for}\; i \in [1,10].$
    The analytic integral is easily found by the difference of the CDF at the prior bounds, and it factorizes over dimension.
    NS using the analytic Gaussian PDF and a \coppuccino{} flow trained on target samples both recover the analytic $\log \mathcal{Z}$ (dashed).
    The flow likelihood introduces no detectable evidence bias.
  }
  \label{fig:gaussian-evidence}
\end{figure}

With the analytic evidence in hand, we turn to nested sampling (Fig.~\ref{fig:gaussian-evidence}).
Our first test case uses the Gaussian PDF directly as the likelihood.
The nested sampler returns a mean and standard deviation of the log evidence, and it recovers the analytic value within one standard deviation.
Next, we use the trained \coppuccino{} flow as the likelihood in the NS.
Again we recover the analytically integrated evidence within one standard deviation, and the result is unbiased relative to the Gaussian-PDF test case.
Taken together, these tests show that our method introduces no inherent bias or loss of accuracy.

\subsection{Simulation study design}\label{sec:methods:sims}

\begin{figure}
  \centering
  \begin{tikzpicture}[
    font=\sffamily\small,
    node distance=9mm and 12mm,
    >={Stealth[length=2.2mm,width=2mm]},
    every edge/.style={draw,thick,->},
    base/.style   ={rectangle, draw, thick, align=center,
                    text width=26mm, minimum height=11mm},
    data/.style   ={base, rounded corners=2pt, fill=blue!8},
    proc/.style   ={base, fill=orange!15},
    flow/.style   ={base, double, double distance=0.8pt, fill=green!12},
    result/.style ={base, rounded corners=2pt, fill=red!10},
  ]

  \node[data]               (priors) {Sample model priors and target SNR $\in [3, 8]$};
  \node[proc, right=of priors] (sim)  {Simulate PTA\\\footnotesize iterate towards target SNR};
  \node[proc, below=of sim] (mcmc)    {Free-spectrum MCMC\\\footnotesize NUTS};
  \node[flow, left=of mcmc] (flow)    {Train \mbox{\textsc{coppuccino}} flow\\\footnotesize $p(\boldsymbol{\rho}\mid \boldsymbol{\delta t})$};
  \node[proc, below=of flow] (ns)     {Nested sampling\\\footnotesize using flow \\\footnotesize evidence $\log \mathcal{Z}_M$};
  \node[result, right=of ns] (bf)     {Bayes factor\\\footnotesize $\log \mathrm{BF}^{\mathcal{M}_1}_{\mathcal{M}_2}=\log \mathcal{Z}_{\mathcal{M}_1}-\log \mathcal{Z}_{\mathcal{M}_2}$};

  \draw (priors) edge (sim);
  \draw (sim)    edge (mcmc);
  \draw (mcmc)   edge (flow);
  \draw (flow)   edge (ns);
  \draw (ns)     edge (bf);

  \draw[->, thick, dashed]
    ($(ns.west) + (0,3mm)$) .. controls +(left:6mm) and +(left:6mm) ..
    ($(ns.west) + (0,-3mm)$)
    node[midway, left=0.5mm, font=\scriptsize\itshape, align=center] {$\times\,2$};

\end{tikzpicture}
  \caption{
    Flow chart showing simulation study pipeline.
    We begin by drawing a target SNR, sampling from the models' prior spaces, and injecting a signal using those parameters.
    The SNR of the injected signal most likely will not match the target, so we iterate on the signal injection as described in Section \ref{sec:tuning-snr}.
    After achieving the desired SNR, we run a PTA free-spectrum analysis and train a \coppuccino{} flow on the resulting posterior.
    This flow is then used in the \blackjaxns{} nested sampler to produce Bayesian evidences used for model comparison.
  }
  \label{fig:pipeline}
\end{figure}
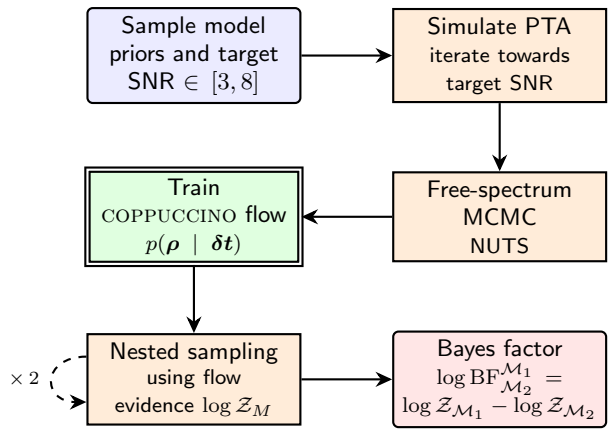

Fig.~\ref{fig:pipeline} summarizes the full simulation study, which we now describe in detail.

\subsubsection{Choosing injection parameters}
We begin by sampling parameters from the prior space of two models: a power law and a log-normal (which we will refer to as ``Gaussian'').
The PSDs of these models in units of $\unit{\second\cubed}$ are
\begin{align}
  \label{eq:10}
  \mathrm{PSD}_{i,\; \mathrm{PL}} &= \frac{A^2}{12 \pi ^2f_{yr}^{3}}\left(\frac{f_i}{f_{yr}}\right)^{-\gamma},\;\mathrm{and}\\
  \mathrm{PSD}_{i,\; \mathrm{Gauss}} &= \frac{A^2}{12 \pi ^2f_{yr}^{3}} \exp\left[\frac{-\left(\log{\left(\frac{f_i}{f_{\mathrm{peak}}}\right)}\right)^2}{ 2 \sigma^2}\right],
\end{align}
where the subscript denotes evaluation at the $i^{\mathrm{th}}$ frequency.
These models are specifically chosen as representatives of two signal classes: signals of astrophysical or cosmological origin.
The power law is the benchmark PSD for a GWB produced by a collection of supermassive black hole binaries (SMBHBs) \cite{phinneyPracticalTheoremGravitational2001}.
This is a simple model, but it is the one used in all GWB ``detection'' pipelines.
The log-normal represents a GWB sourced by scalar-induced gravitational waves (SIGWs).
SIGWs are produced by modified scalar power spectra that result from modified expansion histories in the early universe \citep{domenechScalarInducedGravitational2021,Pi:2020otn}.

As seen in Table \ref{tab:param-dist}, the ``injection'' prior space is more narrow than the ``recovery''.
We deliberately made this choice to keep the models near regimes where they are genuinely distinguishable.
Our study's purpose is to prove the efficacy of our method when we know the models are separable.
It is not meant to explore the areas of parameter space where the models can emulate one another and become indistinguishable.
A study using parameters in those regimes does not strictly test the capabilities of our new method: even a perfect classifier would produce order unity Bayes factors if the models are degenerate with each other (modulo Occam penalty).\footnote{The question of ``when is a cosmological model detectable over a power law'' is nonetheless an interesting and important one. It will be explored in an upcoming work by \citet{cosmoforecast}.}

\begin{table}
  \centering
\begin{tabular}{lll}
  \toprule
  Parameter & Injection prior & Recovery prior \\
  \midrule
  $\logten A_{PL}$ & Uniform(-15.0, -13.5) & Uniform(-18, -11)  \\
  $\gamma_{PL}$ & Uniform(2.0, 6.0) & Uniform(0.0, 7.0)  \\
  $\logten A_{\textrm{Gauss}}$ & Uniform(-14.0, -10.5) & Uniform(-17, -10)  \\
  $\logten f_{\textrm{peak, Gauss}}$ & Uniform(-8.45, -8.05) & Uniform(-8.55, -8.0)  \\
  $\sigma_{\textrm{Gauss}}$ & Uniform(0.3, 0.6) & Uniform(0.3, 1.0)  \\
  \bottomrule
\end{tabular}
\caption{
  Table of parameters used to build simulated PTA datasets and run inferences.
  The injection and recovery priors are not identical because we wish to simulate signals that are \emph{distinct}.
  There are pathological regions of parameter space where these models are indistinguishable, \textit{e.g.} \(\sigma_\mathrm{Gauss} \gg 1\).
}
\label{tab:param-dist}
\end{table}

\subsubsection{Tuning the SNR}\label{sec:tuning-snr}
After sampling model parameters from the injection priors, we also pull a target SNR $\in [3,8]$ that mimics the range of current PTA SNRs \citep{agazieComparingRecentPulsar2024}.
The model parameters that we sampled are \emph{highly} unlikely to match that target SNR, so we iteratively move towards the target SNR by resampling the amplitude parameter only.

Assume we first sample an amplitude parameter $A_0$ that gives $\mathrm{SNR}_0$ in combination with the other GWB parameters and constant noise.
The SNR is approximately
\begin{equation}
  \label{eq:1}
  \mathrm{SNR}_0 \approx A_0 / \mathrm{noise},
\end{equation}
for a given signal amplitude $A$.
For our sampled parameters, there is an $A_1$ that gives the target $\mathrm{SNR}_1$.
Then, constructing the same ratio as Eq.~(\ref{eq:1}) and equating, we can solve for $A_1$ in terms of $A_0$ and the two SNRs:
\begin{equation}
  \label{eq:2}
  A_1 \approx \frac{\mathrm{SNR}_1}{\mathrm{SNR}_0} A_0.
\end{equation}

However, this relationship is not perfect, and we will most likely not reach the target SNR in one go.
Therefore, we iterate this relationship a number of times, replacing $A_0$ and $\mathrm{SNR}_0$  with the current amplitude and SNR each time.
Once our current SNR is within 0.5 of the target SNR or the relative error is less than 5\%, we cease iterating and save the amplitude value.

\subsubsection{Using the likelihood as a generative model}
Next, we simulate timing residuals by treating the likelihood as a generative model.
Once we have defined the probability of the data given parameters (the likelihood), we can sample data from that distribution conditioned on the parameters. 

We build an HD-correlated likelihood in \discovery{} mimicking that of the NANOGrav 15yr flagship analysis \cite{NANOGrav:2023gor}.
The likelihood setup requires times of arrival, timing model design matrices, and pulsar sky positions, which PTA codes have begun storing in the Apache Feather file format \cite{richardsonArrowIntegrationApache2026}.
The \discovery{} GitHub repository \cite{vallisneriNanogravDiscovery0512025} contains the NG15yr data \cite{NANOGrav:2023hde} stored in this format, and we use this data as the basis of our simulations.

Given the data in this format, we construct a PTA likelihood and use it to simulate timing residuals.
This is possible because the PTA likelihood is a Gaussian likelihood on the residuals with covariance given by our modeled processes.
To simulate residuals, we sample from this zero-mean, multivariate Gaussian where the covariance matrix is constructed from the sampled GWB parameters and maximum \textit{a posteriori} estimates of the noise parameters.

As mentioned previously, we choose two representative GWB models: a power law and a log-normal.
For each model, we simulate fifty independent realizations following the procedure described above.

\subsubsection{Calculating the Bayesian evidences}
With the simulated data in hand, we produce a flow-based likelihood-surrogate, as described in Section~\ref{sec:methods:flow-accel}, for each simulation.
Then, to acquire Bayesian evidences and parameter posteriors, we run NS with \blackjaxns{}.
The output of \blackjaxns{} gives us the log Bayesian evidence, $\log \mathcal{Z}$, and posterior samples.
We run NS using a power law and a log-normal model for each simulation, giving us 200 NS runs total.
As a cross-check on the evidence calculation, we run NUTS on the full PTA likelihood and estimate the evidence with a learned-harmonic-mean estimator \citep{mcewenMachineLearningAssisted2023}, an independent check that does not use the surrogate.
Then, to ensure that our surrogate returns the same parameter constraints as a typical PTA likelihood, we compare to the NS+surrogate posteriors.

After we have completed our NS runs and cross-checks, all that remains is to calculate the Bayes factors.
For each simulation, we calculate the Bayes factor for power law over the log-normal using the stored log evidences,
\begin{equation}
  \label{eq:3}
  \logten \BFPG = \logten \mathcal{Z}_\mathrm{PL} - \logten \mathcal{Z}_\mathrm{Gauss}.
\end{equation}

\section{Results}\label{sec:results}

\subsection{Flow quality on free-spectrum posteriors}\label{sec:results:flow-quality}
\begin{figure}
  \includegraphics[width=\columnwidth]{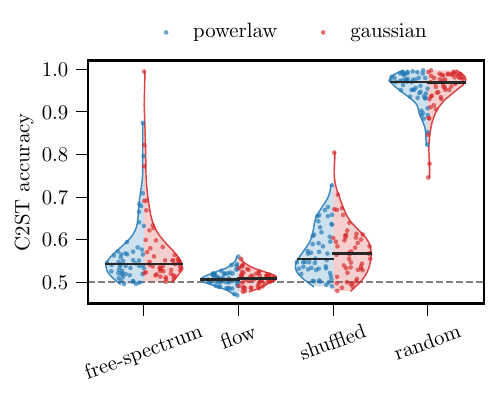}
  \caption{
    Fidelity of the flow to the free-spectrum posterior, measured by a classifier two-sample test (C2ST).
    We train a gradient-boosted classifier to distinguish two sample sets.
    Its accuracy (left axis) is $0.5$ when the sets are indistinguishable and $1.0$ when trivially separable.
    Each split violin is the distribution over all $50$ realizations per model (power law left/blue, Gaussian right/red), overlaid with a sina point cloud with black tick marks showing the median.
    We show four comparisons against the free-spectrum chains: ``free-spectrum'', two independent halves of the same chain; ``flow'', the \coppuccino{} flow against the chain; ``shuffled'', the chain against a copy with each frequency bin independently permuted (marginals preserved, cross-bin correlations destroyed); and ``random'', the chain against a uniform distribution over the matched per-bin support.
    The flow is indistinguishable from the posterior: its violin overlaps the free-spectrum control and contains the $0.5$ ``indistinguishable'' line.
    Destroying correlations or substituting a random distribution is readily detected, confirming both the flow's fidelity and the C2ST's sensitivity to the different distributions.
  }
  \label{fig:c2st}
\end{figure}

Before turning to the model-comparison results, we first ensure that our flows accurately capture the free-spectrum posterior.
We quantify the fidelity with a classifier two-sample test (C2ST)~\citep{lopez-pazRevisitingClassifierTwoSample2018,friedmanMultivariateGoodnessofFitTwoSample2004} with \textsc{scikit-learn}~\citep{scikit-learn}.
We train a gradient-boosted classifier to distinguish flow samples from free-spectrum MCMC samples and report its cross-validated accuracy in Fig.~\ref{fig:c2st}.
The interpretation of the accuracy is as follows: a value of $0.5$ means the two sample sets are indistinguishable, while $1.0$ means they are trivially separable.

We anchor the scale with four comparisons against the free-spectrum chain.
As null test-case, two independent halves of the same free-spectrum chain are only weakly distinguishable (median accuracy $0.54$), reflecting finite-sample noise alone.
The flow sits at this null (median accuracy $0.51$): a classifier cannot separate flow draws from true posterior draws any better than it can separate the posterior from itself.
By contrast, destroying the inter-frequency correlations by independently permuting each bin (\emph{shuffled}) raises the accuracy to $0.57$, and a random distribution over the matched support is trivially separated (\emph{random}, median accuracy $0.97$).
The flow is therefore indistinguishable from the free-spectrum posterior: its accuracy distribution overlaps the \emph{free-spectrum} null and straddles the $0.5$ line.
The \emph{shuffled} control sits only modestly above the null because the test is a insensitive to correlation structure, as noted above, so we examine the cross-bin correlations the flow must preserve directly below.

\begin{figure}
  \includegraphics[width=\columnwidth]{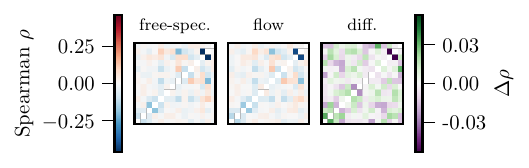}
  \caption{
    Cross-bin correlation structure for a representative power law realization.
    We use Spearman's correlation to capture any \emph{monotonic} correlations.
    The free-spectrum posterior is shown on the left, the flow in the middle, and their difference to the right.
    We mask out the diagonals (they are identity) such that the off diagonal scale is dominant.
    The free-spectrum and flow panels share a color scale.
    The difference is roughly an order of magnitude smaller and is shown on its own expanded scale with a distinct colormap.
    The flow reproduces the joint-correlation structure, not just the marginals.
    Across all realizations the flow matches the chain's off-diagonal Spearman correlations with RMS $0.01$.
  }
  \label{fig:flow-correlations-pl}
\end{figure}

\begin{figure}
  \includegraphics[width=\columnwidth]{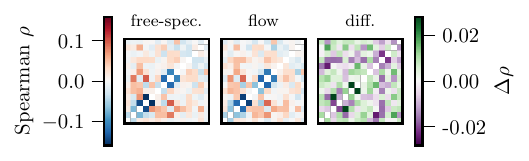}
  \caption{
    Cross-bin correlation structure for a representative Gaussian realization.
    The setup and visualization scheme are the same as Fig.~\ref{fig:flow-correlations-pl}.
    As in Fig.~\ref{fig:flow-correlations-pl}, the RMS of the off-diagonal Spearman correlations is 0.01.
  }
  \label{fig:flow-correlations-g}
\end{figure}

We examine the inter-frequency correlations more directly with a Spearman rank-correlation plot (Fig.~\ref{fig:flow-correlations-pl} and~\ref{fig:flow-correlations-g}), where we mask out the diagonal to make the inter-frequency correlations apparent.
The difference between the flow and free-spectrum-posterior correlations is near zero, with an RMS of $0.01$, showing that the flows accurately recover the inter-frequency correlations.

\subsection{Parameter recovery and calibration}\label{sec:results:recovery}
\begin{figure}
  \includegraphics[width=\columnwidth]{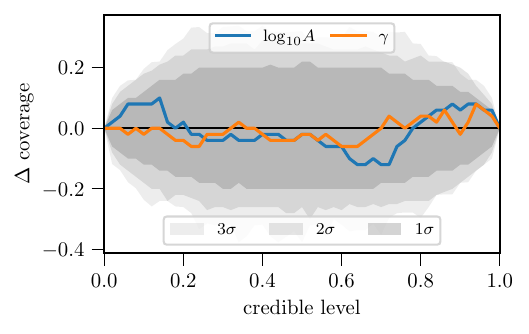}
  \caption{
    Calibration of the flow-accelerated nested sampling against a full-likelihood, traditional MCMC run using the power law model.
    Each line is the difference of the two empirical coverage curves (flow$+$NS $-$ traditional) for one parameter.
    A difference consistent with zero means the surrogate is as calibrated as the traditional inference.
    The shaded bands are simultaneous $1\sigma/2\sigma/3\sigma$ envelopes for the hypothesis that the deviations from zero are Monte Carlo error \citep{sailynojaGraphicalTestDiscrete2022}.
    To construct the bands, we create two populations of samples from $\mathcal{U}(0,1)$, find their ECDFs, and take their difference.
    We repeat this process 2000 times and draw the 1, 2, and 3$\sigma$ regions as the bands wide enough to contain that fraction of the difference curves in their entirety.
  }
  \label{fig:param-recovery-pl}
\end{figure}

\begin{figure}
  \includegraphics[width=\columnwidth]{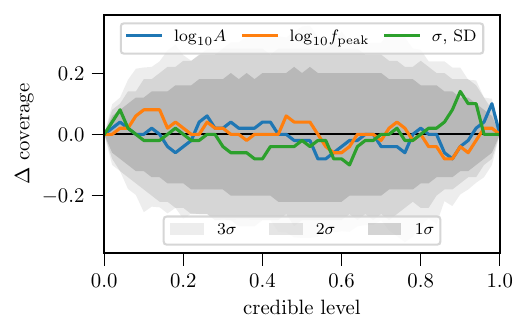}
  \caption{As Fig.~\ref{fig:param-recovery-pl}, for the Gaussian injection parameters.}
  \label{fig:param-recovery-g}
\end{figure}

Next, we show that the flow-accelerated NS recovers the same parameter constraints as a traditional PTA analysis.
To do so, we construct calibration plots for both of our models.
Typically, one would create a probability--probability (P--P) plot, the standard simulation-based-calibration check in which one shows the fraction of times that a parameter is recovered within a given highest-density interval.
Equivalently, this is the empirical CDF of the posterior quantiles (the probability-integral-transform values), which is diagonal when the method is calibrated and the injections are drawn from their priors.
In our case, however, we are comparing two methods recovering the \emph{same} parameters, and we do not draw uniformly from the prior.
We therefore construct a variant of this plot where we show the difference between the curves that the two methods would produce.
If the methods agree on their calibration, this difference should be identically zero.
There is, however, inherent Monte Carlo error in this process.
To show the deviations from zero that could arise entirely from Monte Carlo error, we construct bands following the prescription of~\citet{sailynojaGraphicalTestDiscrete2022}.
These bands mark the one-, two-, and three-sigma regions of deviation from zero consistent with Monte Carlo error alone.
Both calibration figures (Fig.s~\ref{fig:param-recovery-pl} and~\ref{fig:param-recovery-g}) show that the difference lies near zero and entirely within the $1\sigma$ band, implying that our method is identical in calibration to the traditional analysis and that we recover the same posteriors.

\subsection{Aggregate model selection}\label{sec:results:model-selection}
Finally, we turn our attention to the results of the model comparison itself.
For each simulation we have a log-evidence for both models, giving us 200 log-evidences in total.
From each pair, we form the Bayes factor in favor of the power law: $\logten\BFPG$ (Eq.~\ref{eq:3}).
This should be positive for the power-law (``PL'') injections and negative for the log-normal (``Gauss'') injections.

\begin{figure}
  \includegraphics[width=\columnwidth]{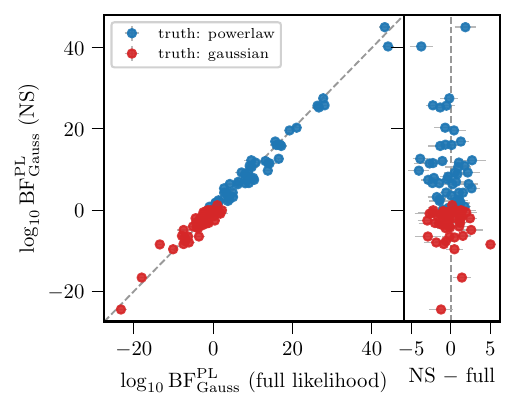}
  \caption{NS $\logten\BFPG$ versus an independent learned-harmonic-mean estimate on NUTS-sampled posteriors, with residuals in the right-hand panel.
  The difference between the NS results and the learned harmonic mean have a mean of -0.18 and standard deviation of 1.62.}
  \label{fig:ns-harmonic}
\end{figure}

To cross-check our results, we compare against an evidence estimate built from the full PTA likelihood, bypassing the surrogate entirely.
For each simulation we run NUTS on the full PTA likelihood for both test models.
Then, we estimate each model's evidence from those posterior samples with a learned harmonic-mean estimator \citep{mcewenMachineLearningAssisted2023}, using a temperature-scaled Gaussian as the internal importance density.
Because this uses the full likelihood rather than our flow surrogate, it is an independent check of our NS Bayes factors.
We plot the two against each other in Fig.~\ref{fig:ns-harmonic}.
If they agree, they should lie along the diagonal.
That is indeed what we see, confirming that the Bayes factors recovered from our surrogate-based NS are in agreement with those of the full PTA likelihood.
Moreover, in Section~\ref{sec:results:recovery} we show that our surrogate recovers the same posteriors as the full PTA likelihood.
The surrogate Bayes factor also equals the full-likelihood Bayes factor, which we show analytically in Section~\ref{sec:methods:marginalization}.
The only limit on this equality is the fidelity of the flow, which we test in Section~\ref{sec:results:flow-quality}.

\begin{figure}
  \includegraphics[width=\columnwidth]{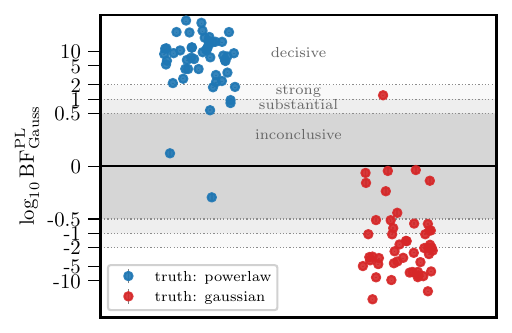}
  \caption{
    Per-realization $\logten\BFPG$ grouped by the true, injected model.
    Shaded bands are the Jeffreys' scale: inconclusive, substantial, strong, and (unshaded) decisive.
    Positive values favor power law, negative favor the Gaussian.
    The bulk of the Bayes factors lie in the ``substantial'' region and beyond.
    The outliers at lower significance (or with incorrect sign) come from injections near the boundaries of the injected parameter space, where the models can begin to become confused.
  }
  \label{fig:bf-results}
\end{figure}

We also plot the full distribution of Bayes factors with Jeffreys' scale overlaid in Fig.~\ref{fig:bf-results}.
The correct model is predominantly identified in the substantial and decisive regimes, with a small number of outliers at lower significance.
These outliers are mostly due to injections at the boundaries of the prior space, where the two models begin to become confused.

The magnitudes of our Bayes factors, while large, are genuine rather than an artifact of our surrogate.
The independent full-likelihood harmonic-mean estimate of Fig.~\ref{fig:ns-harmonic}, which never touches our tail model, reproduces them.
Our large Bayes factors are caused by our models routinely predicting power in the tails of our data.
Correctly recovering Bayes factors of this magnitude requires our tail treatment (Sec.~\ref{sec:tails}).
Without it, the previous prescription of holding the surrogate likelihood constant beyond the training support \citep{lambNeedSpeedRapid2023} results in maximum Bayes factors near $\logten\BFPG \sim 5$, severely under-penalizing disfavored models.

\begin{figure}
  \includegraphics[width=\columnwidth]{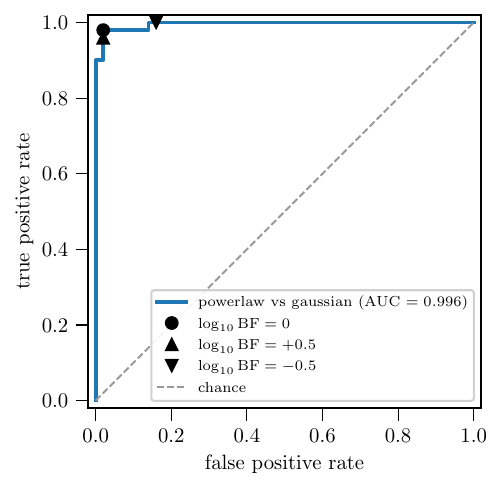}
  \caption{
    Model discrimination from a threshold on $\logten\BFPG$.
    We treat the Bayes factor as a binary classifier with power law injections as the positive class.
    The natural decision point $\logten\BFPG=0$ and Bayes factors corresponding to ``substantial'' on the Jeffreys' scale are marked.
    The area under the curve (AUC) score is 0.996, showing that the Bayes factors produced by our methods are excellent classifiers.
  }
  \label{fig:roc}
\end{figure}

To quantify the quality of our Bayes factors as model discriminators, we construct a receiver operating characteristic (ROC) curve and the associated area under the curve (AUC) in Fig.~\ref{fig:roc}.
We treat the Bayes factor as a binary classifier, taking the power-law injections as the positive class and sweeping a decision threshold across $\logten\BFPG$.
For each threshold the ROC traces the true-positive rate against the false-positive rate.
A perfect classifier hugs the left and upper edges of the plot, giving an AUC of unity, whereas a random classifier follows the diagonal with an AUC of $0.5$.
We recover a ROC curve confined to the upper-left corner and an AUC of $0.996$.
Together, these show that our recovered Bayes factors are excellent model discriminators.

\section{Conclusions}\label{sec:conclusions}
We present a fast, computationally-inexpensive method for Bayesian model comparison of PTA data.
Our method uses a normalizing-flow likelihood surrogate that, in tandem with the \blackjaxns{} nested sampler, can produce parameter posteriors and Bayesian evidences in minutes on consumer hardware.
Compared to traditional PTA analyses, this flow-based NS approach yields unbiased parameter posteriors (see Fig.s \ref{fig:param-recovery-pl} and \ref{fig:param-recovery-g}).
We validate the quality of the calculated Bayesian evidences as model comparison metrics using a receiver operating characteristic (ROC) and the associated area under the curve (AUC) score in Fig. \ref{fig:roc}.
The AUC score is 0.996, proving that the Bayesian evidences calculated with our new method are reliable for model comparison.

Our method is general and flexible.
It need not be used with a nested sampler: the likelihood surrogate can be supplied to any sampler.
Furthermore, it is \emph{automatically differentiable} via \textsc{JAX}, so it can be used in Hamiltonian Monte Carlo, maximum \textit{a posteriori} parameter estimates via gradient descent, and similar gradient-based algorithms.

A future avenue we intend to pursue is training our likelihood surrogate directly on the Fourier coefficients instead of on their summary statistics (the variance parameters $\bs{\rho}$).
This approach is now feasible due to recent developments that lower the computational cost of sampling the Fourier coefficients \citep{gundersenNewFrameworkLightningfast2026}.
Our refitting scheme applied to the Fourier coefficients would then also work for continuous waves (CW) from loud, nearby SMBHBs that are discernible above the gravitational-wave background.
Training on the coefficients additionally enables the use of \emph{non-Gaussian} signals, which we expect in general from realistic populations of SMBHBs and some cosmological models \cite{Lamb:2024gbh,Lamb:2025niq,kumarNonGaussianStochasticGravitational2021}.
In addition, because our method learns inter-frequency correlations, it can be easily applied to posteriors produced by PTA likelihoods that include these correlations \citep{crisostomiDiagonalApproximationsImproved2025}.

The code and all data products produced as part of this study are publicly available.
We intend to integrate our method with existing PTA analysis codes and provide trained normalizing flows for current PTA datasets. 

As we enter the detection era of PTA science, the main goal evolves from the detection of the GWB to its characterization and source identification.
Our method enables and democratizes the means to reliably accomplish these goals.

\begin{acknowledgments}
The authors are funded as part of the NANOGrav Collaboration through the National Science Foundation (NSF) NANOGrav Physics Frontiers Center award \#2020265. JSH acknowledges support from NSF CAREER Award No. 2339728. JSH is also supported through an Oregon State University start up fund.
\end{acknowledgments}

\section*{Data availability}\label{sec:conclusions:code}
\coppuccino{} is available at \url{https://github.com/AaronDJohnson/coppuccino} or on PyPI; the analysis pipeline used in this work is available at \url{https://github.com/davecwright3/fast-and-flowrious}.

\bibliography{refs}

\end{document}